\documentstyle[graphicx, 12pt]{article}
\begin{document}

\small
\hoffset=-1truecm
\voffset=-2truecm
\title{\bf The calculation of the thermodynamic quantities of the Bardeen black hole}
\author{Jingyun Man \hspace {1cm} Hongbo Cheng\footnote {E-mail address: hbcheng@ecust.edu.cn}\\
Department of Physics, East China University of Science and
Technology,\\ Shanghai 200237, China\\
The Shanghai Key Laboratory of Astrophysics, Shanghai 200234,
China}

\date{}
\maketitle

\begin{abstract}
In this work we research on the thermodynamical properties of the
Bardeen black holes. We compute the series of new thermodynamic
quantities such as local temperature, heat capacity, off-shell
free energy of this kind of black hole in detail. We further
analyze the thermodynamical characteristics of the Bardeen black
hole by varying its charge $q$ to check the existence and
stability of the black hole.
\end{abstract}
\vspace{7cm} \hspace{1cm} PACS number(s): 04.70.Bw, 14.80.Hv\\
Keywords: Black hole; Thermodynamics; Phase transition

\newpage

\noindent \textbf{I.\hspace{0.4cm}Introduction}

The thermodynamics of various kinds of black holes which are space
regions that nothing can escape from become a focus recently. More
than thirty years ago, Bekenstein predicted that black holes
should have finite temperature which is not equal to zero and
further the entropy of a black hole is proportional to its surface
area [1-3]. Hawking et. al. researched on the quantum mechanics of
scalar particles around a black hole to reveal that the black hole
has a thermal radiation with the temperature due to its surface
gravity [4-6]. A lot of notes on the phase transitions of black
holes in the frame of semiclassical gravity were listed in Ref.
[7]. The thermodynamic characteristics of modified Schwarzschild
black hole and quantum-corrected Schwarzschild black hole have
been investigated [8-10]. The thermodynamic behaviours including
phase transition in Born-Infeld-anti-de Sitter black holes were
probed in virtue of various ways [11, 12].

A kind of black hole whose distinct feature is that its spacetime
is regular without a singularity but a horizon was put forward by
Bardeen [13]. Some other regular black holes were also discussed
[14, 15]. More efforts have been made for the Bardeen black holes.
A Bardeen black hole and some similar regular black holes as exact
solutions to a model of nonlinear electrodynamics coupled to
Einstein gravity was studied [16-19]. Eiroa and Sendra examined
the gravitational lensing of the Bardeen black holes [20]. The
geodesic motion of the test particles around the regular black
hole was also studied [21]. The gravitational and electromagnetic
stability of this kind of black holes have been explored [22]. The
quasinormal modes of the scalar field perturbations of the Bardeen
black hole is discussed [23].

It is necessary to scrutinize the thermodynamic properties of the
Bardeen black holes. The quantum corrections to some
thermodynamical quantities such as temperature and entropy for a
Bardeen black hole were evaluated within a quantum tunneling
approach over  semiclassical approximation [24]. In the
noncommutative spacetime the temperature, entropy, singularity,
horizon and mass function of the charged regular black hole are
examined [25]. In Ref. [24] and [25], only the quantum and
noncommutative corrections to the Hawking temperature and entropy
of the Bardeen black hole were revealed, and the further research
on the thermodynamic characteristics of this kind of black hole
have not been performed. We wish to find the other thermodynamical
characteristics of the Bardeen black hole. We plan to derive and
calculate the local temperature, heat capacity and off-shell free
energy of the Bardeen black hole in order to check the existence
and stability of the black hole. We certainly wonder how the
thermodynamical quantities related to the magnetic charge of the
black hole. The results and conclusions will be listed finally.

\vspace{0.8cm} \noindent \textbf{II.\hspace{0.4cm}The
thermodynamics quantities Bardeen black holes}

The regular static-charged black hole named as Bardeen black hole
is the gravitational field of a magnetic monopole arising from
nonlinear electrodynamics [13, 16]. According to the Einstein
equations and field equations from the proposed action involving
the nonlinear electrodynamic term, the metric of the so-called
Bardeen black hole as a static spherically solution for the
equations is given as,

\begin{equation}
ds^{2}=f(r)dt^{2}-\frac{dr^{2}}{f(r)}-r^{2}(d\theta^{2}+\sin^{2}\theta
d\varphi^{2})
\end{equation}

\noindent where

\begin{equation}
f(r)=1-\frac{2Mr^{2}}{(r^{2}+q^{2})^{\frac{3}{2}}}
\end{equation}

\noindent Here $q$ and $M$ are the magnetic charge and the mass of
the magnetic monopole respectively. As $q=0$, the spacetime
structure of Bardeen black hole certainly recover to be the
Schwarzschild metric.

Asymptotically, the metric function $f(r)$ can be approximated as,

\begin{equation}
f(r)\approx1-\frac{2M}{r}+\frac{3Mq^{2}}{r^{3}}+O(\frac{1}{r^{5}})
\end{equation}

\noindent which is different from Reissner-Nordstrom metric.
According to Figure 1, if the horizons exist, the inner one as
Killing horizon and the outer one as event horizon, the charge $q$
has to be smaller than $\frac{4\sqrt{3}}{9}M$. When
$q=\frac{4\sqrt{3}}{9}M$, the two horizons meet. When
$q>\frac{4\sqrt{3}}{9}M$, there is no horizon for $f(r)$.

The event horizon of metric (1) is located at,

\begin{eqnarray}
r_{H}=[\frac{1}{3}(4M^{2}-3q^{2})+\frac{\sqrt[3]{2}}{3}
(32M^{6}-72M^{4}q^{2}+27M^{2}q^{4}+3\sqrt{3}
\sqrt{27M^{4}q^{^{8}-16M^{6}q^{6}}})^{\frac{1}{3}}\nonumber\\
-\frac{24M^{2}q^{2}-16M^{4}}{3\sqrt[3]{2}(32M^{6}
-72M^{4}q^{2}+27M^{2}q^{4}+3\sqrt{3}\sqrt{27M^{4}q^{8}
-16M^{6}q^{6}})^{\frac{1}{3}}}]^{\frac{1}{2}}\hspace{1cm}
\end{eqnarray}

\noindent The relation between the total mass and the event
horizon also results from metric (1),

\begin{equation}
M=\frac{(r_{H}^{2}+q^{2})^{\frac{3}{2}}}{2r_{H}^{2}}
\end{equation}

\noindent then the minimal mass $M_{0}$ reads,

\begin{equation}
M_{0}=\frac{3\sqrt{3}}{4}q
\end{equation}

\noindent which has something to do with the magnetic charge. The
total mass as a function of event horizon and charge is depicted
in Figure 2. The Hawking temperature can be calculated as,

\begin{eqnarray}
T_{H}(q)=\frac{1}{4\pi}[\sqrt{-g^{tt}g^{rr}}g'_{tt}]|_{r=r_{H}}\nonumber\\
=\frac{1}{4\pi}\frac{r_{H}^{2}-2q^{2}}{r_{H}(r_{H}^{2}+q^{2})}\hspace{2cm}
\end{eqnarray}

\noindent where the prime stands for the derivative with respect
to the radial coordinate $r$. According to Ref. [26], the local
temperature is given by,

\begin{eqnarray}
T_{loc}(q)=\frac{T_{H}(q)}{\sqrt{f(r)}}\hspace{3cm}\nonumber\\
=\frac{1}{4\pi}\frac{r_{H}^{2}-2q^{2}}{r_{H}(r_{H}^{2}+q^{2})}
\frac{1}{\sqrt{1-\frac{r^{2}}{r_{H}^{2}}(\frac{r_{H}^{2}+q^{2}}{r^{2}+q^{2}})
^{\frac{3}{2}}}}
\end{eqnarray}

\noindent The temperature associated with the black hole radius
and charge is plotted in Figure 3. It is clear that the shape of
temperature is similar to those of the noncommutative
Schwarzschild black hole and the Reissner-Nordstrom black hole [8]
although the expressions are different. There is a zero for the
local temperature $T_{loc}(q)|_{r_{H}=r_{0}}=0$ and here
$r_{0}=\sqrt{2}q$. Minimizing the local temperature (8) with
respect to the black hole radius $r_{H}$ like $(\frac{\partial
T_{loc}}{\partial r_{H}})_{r}=0$, we find the following equation,

\begin{equation}
(r_{H}^{2}-2q^{2})^{2}+[1-\frac{r^{2}}{r_{H}^{2}}
(\frac{r_{H}^{2}+q^{2}}{r^{2}+q^{2}})^{\frac{3}{2}}]
(18r_{H}^{2}q^{2}-3r_{H}^{4})=0
\end{equation}

\noindent whose real roots are $r_{1}$ and $r_{2}$. The numerical
calculation shows that the smaller radius $r_{1}$ becomes larger
while the larger one $r_{2}$ gets smaller with increasing charge
$q$. The local temperature has two extrema denoted as
$T_{i}=T|_{r_{H}=r_{i}}$ with $i=1, 2$, where $r_{1}$ and $r_{2}$
are the positions of the two extrema respectively. We choose
charge $q=0.5$ to obtain $T_{1}=0.0411$ with $r_{1}=1.385$ and
$T_{2}=0.0204$ with $r_{2}=6.594$. It is also interesting that the
higher temperature $T_{1}$ decreases remarkably and the lower one
$T_{2}$ just increase a little when the charge $q$ becomes larger.
There is one small black hole for $0<T<T_{2}$ and one large black
hole for $T>T_{1}$. When the local temperature is intermediate
like $T_{2}<T<T_{1}$, there will exist three black holes.

According to Bekenstein's approach, the entropy is portional to
the area of event horizon and denoted as [1-3],

\begin{eqnarray}
S=\frac{A}{4}\nonumber\\
=\pi r_{H}^{2}
\end{eqnarray}

\noindent leading $dS=2\pi r_{H}dr_{H}$.

We combine the local temperature (8), entropy (10) and the first
law of thermodynamics $dE_{loc}=T_{loc}dS$ to find that,

\begin{equation}
dE_{loc}=\frac{1}{2}\frac{r_{H}^{2}-2q^{2}}{r_{H}^{2}+q^{2}}
\frac{1}{\sqrt{1-\frac{r^{2}}{r_{H}^{2}}(\frac{r_{H}^{2}+q^{2}}
{r^{2}+q^{2}})^{\frac{3}{2}}}}dr_{H}
\end{equation}

In order to check the stability of the Bardeen black hole, we
should derive the black hole's capacity as,

\begin{eqnarray}
C(q)=(\frac{\partial E_{loc}}{\partial T_{loc}})_{r}\hspace{2.5cm}\nonumber\\
=\frac{4\pi r_{H}^{2}(r_{H}^{2}-2q^{2})(r_{H}^{2}+q^{2})}
{18q^{2}r_{H}^{2}-3r_{H}^{4}+\frac{(r_{H}^{2}-2q^{2})^{2}}
{1-\frac{r^{2}}{r_{H}^{2}}(\frac{r_{H}^{2}+q^{2}}{r^{2}+q^{2}})^{\frac{3}{2}}}}
\end{eqnarray}

\noindent The curves of the heat capacity of Bardeen black hole is
exhibited in Figure 4. The heat capacity is positive for
$r_{0}<r_{H}<r_{1}$ and $r_{H}>r_{2}$, which means that the small
and the large Bardeen black hole can survive for long time. Within
the region $r_{1}<r_{H}<r_{2}$, the black hole heat capacity is
negative, showing that the intermediate black holes decay quickly.
As mentioned above, the positions $r_{1}$ and $r_{2}$ approach
each other as the black hole charge becomes stronger. There are
two zeros for the heat capacity denoted as $C(q)|_{r_{H}=r_{0},
r}=0$. It should be pointed out that the stability of the black
holes can be displayed in their off-shell free energy.

The off-shell free energy is defined as,

\begin{equation}
F_{off}=E_{loc}-TS
\end{equation}

\noindent where $T$ is an arbitrary temperature. According to Eq.
(10) and (11), the off-shell free energy of Bardeen black holes
can be represented in the integral form as,

\begin{equation}
F_{off}(q)=\frac{1}{2}\int_{r_{0}}^{r_{H}}\frac{x^{2}-2q^{2}}
{x^{2}+q^{2}}\frac{1}{\sqrt{1-\frac{r^{2}}{x^{2}}
(\frac{x^{2}+q^{2}}{r^{2}+q^{2}})^{\frac{3}{2}}}}dx-\pi r_{H}^{2}T
\end{equation}

\noindent It is evident that the off-shell free energy expression
is not explicit, but the relation between the off-shell free
energy and the event horizon for several temperatures within the
region $r_{H}\leq r$ can be shown graphically. How the off-shell
free energy of Bardeen black holes changes with the horizon under
several temperatures is exhibited in Figure 5. When $T<T_{2}$, the
curves of off-shell free energy have only one minimum
respectively, which representing a small stable black hole. As the
temperature rises to the value between $T_{2}$ and $T_{1}$, three
black holes will emerge. Among the three black holes, the small
one and the large one are stable while the intermediate black hole
will be unstable, corresponding to two minimum and one maxima of
the off-shell free energy, which is exhibited in Figure 6. If the
temperature is sufficiently high like $T>T_{1}$, there will be
only one minimum for the free energy at the position that the
horizon is larger, and certainly one black hole which is large and
stable will appear.

\vspace{0.8cm} \noindent \textbf{III.\hspace{0.4cm}Discussion}

In this paper we explore the existence and the stability of the
Bardeen black hole in a new direction. We calculate the
thermodynamic quantities such as local temperature, heat capacity,
off-shell free energy for black hole to show how the quantities
change with the horizon $r_{H}$ and the charge $q$. In general,
one small black hole will generate when the temperature is not too
low. With the higher and higher temperature there will be three
black holes. For the sufficiently high temperature, one larger
black hole remains. According to the heat capacity, only when the
event horizon is between $r_{0}$ and $r_{1}$ or larger than
$r_{2}$, the Bardeen black holes can be stable because of the
positive parts of the quantity. Within the region that the size of
the Bardeen black hole is between $r_{1}$ and $r_{2}$, the heat
capacity is negative, corresponding to an unstable black hole. It
can be checked as a requirement for off-shell free energy owing to
its extrema. It is obvious that only when the temperature is
higher than $T_{2}$ but smaller than $T_{1}$ the stable small and
large black holes corresponding to two minima of the off-shell
free energy survive for long time while the middle one due to the
maximum of the free energy will decay quickly. For $T>T_{1}$, one
large Bardeen black hole exists stably. It should be pointed out
that the influence from the charge of black hole on the black hole
size and the critical temperatures is manifest and distinct.

\vspace{1cm}
\noindent \textbf{Acknowledge}

This work is supported by NSFC No. 10875043 and is partly
supported by the Shanghai Research Foundation No. 07dz22020.

\newpage
\begin{figure}
\setlength{\belowcaptionskip}{10pt} \centering
\includegraphics[width=15cm]{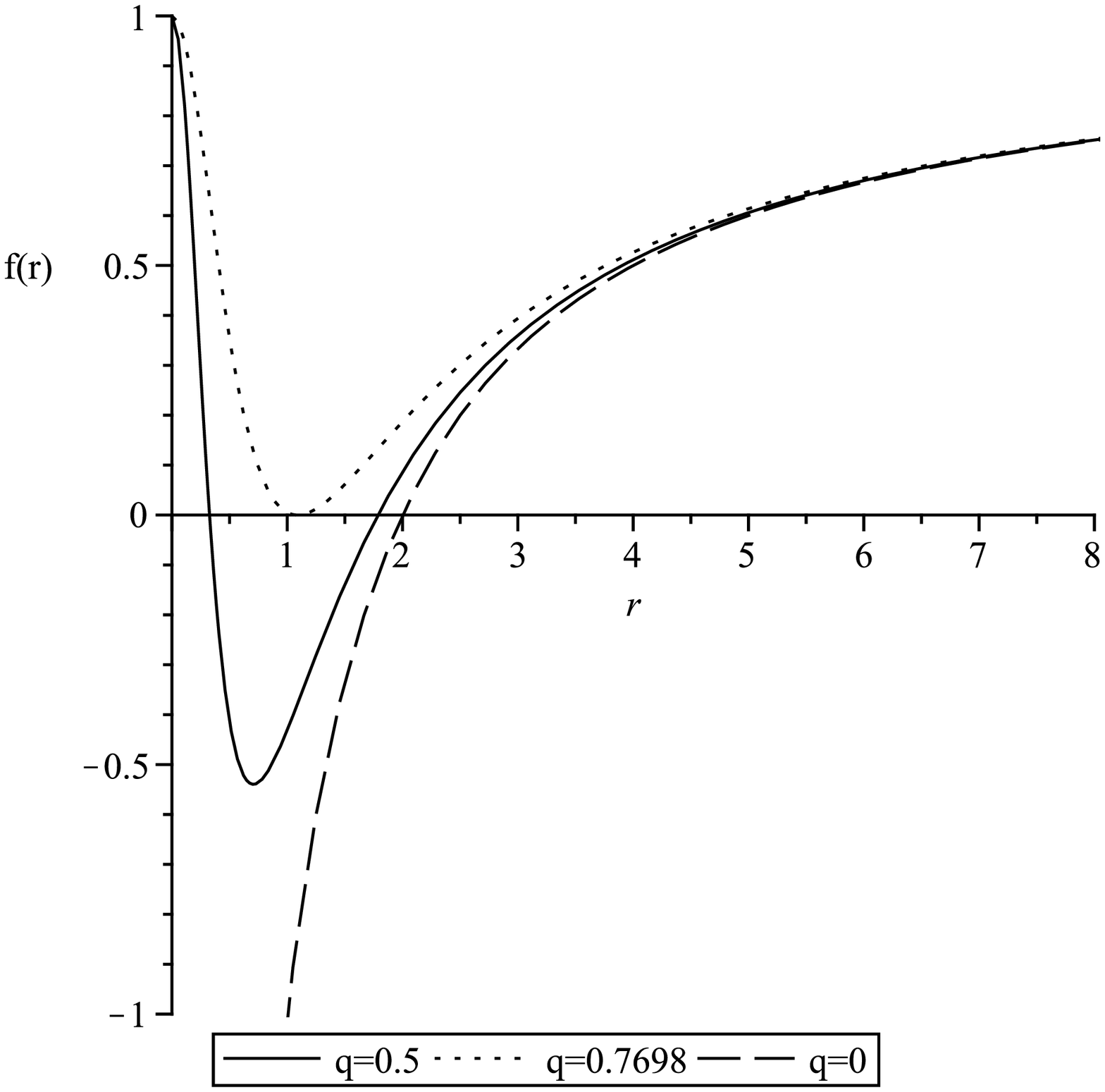}
\caption{The relation between $f(r)$ and $r$ for magnetic charge
$q=0, 0.5, 0.7698$ respectively.}
\end{figure}

\newpage
\begin{figure}
\setlength{\belowcaptionskip}{10pt} \centering
\includegraphics[width=15cm]{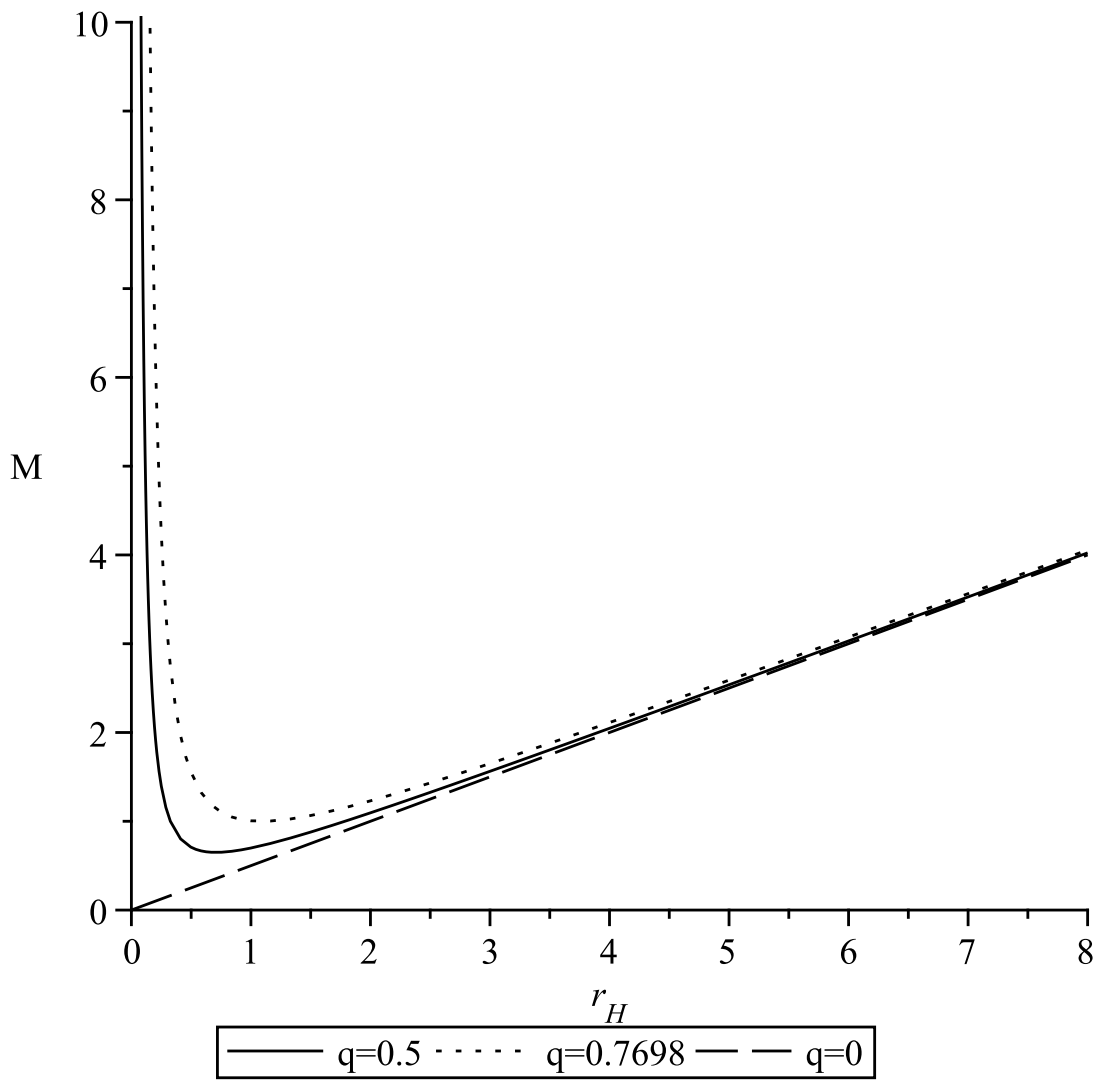}
\caption{The total mass of Bardeen black hole as a function of
event horizon for magnetic charge $q=0, 0.5, 0.7698$
respectively.}
\end{figure}

\newpage
\begin{figure}
\setlength{\belowcaptionskip}{10pt} \centering
  \includegraphics[width=15cm]{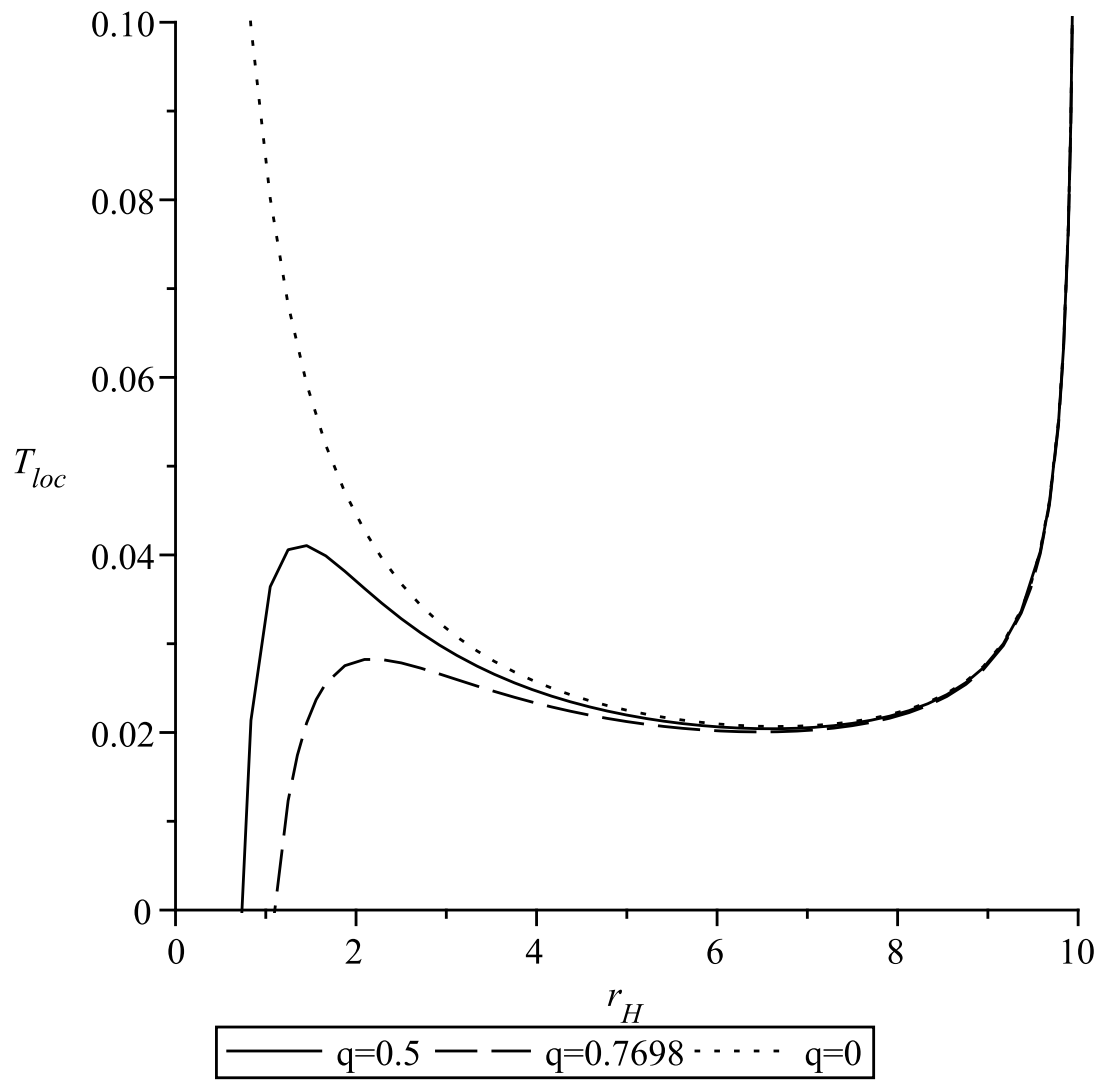}
  \caption{The dependence of local temperature on the event horizon
  for magnetic charge $q=0, 0.5, 0.7698$ respectively.}
\end{figure}

\newpage
\begin{figure}
\setlength{\belowcaptionskip}{10pt} \centering
  \includegraphics[width=15cm]{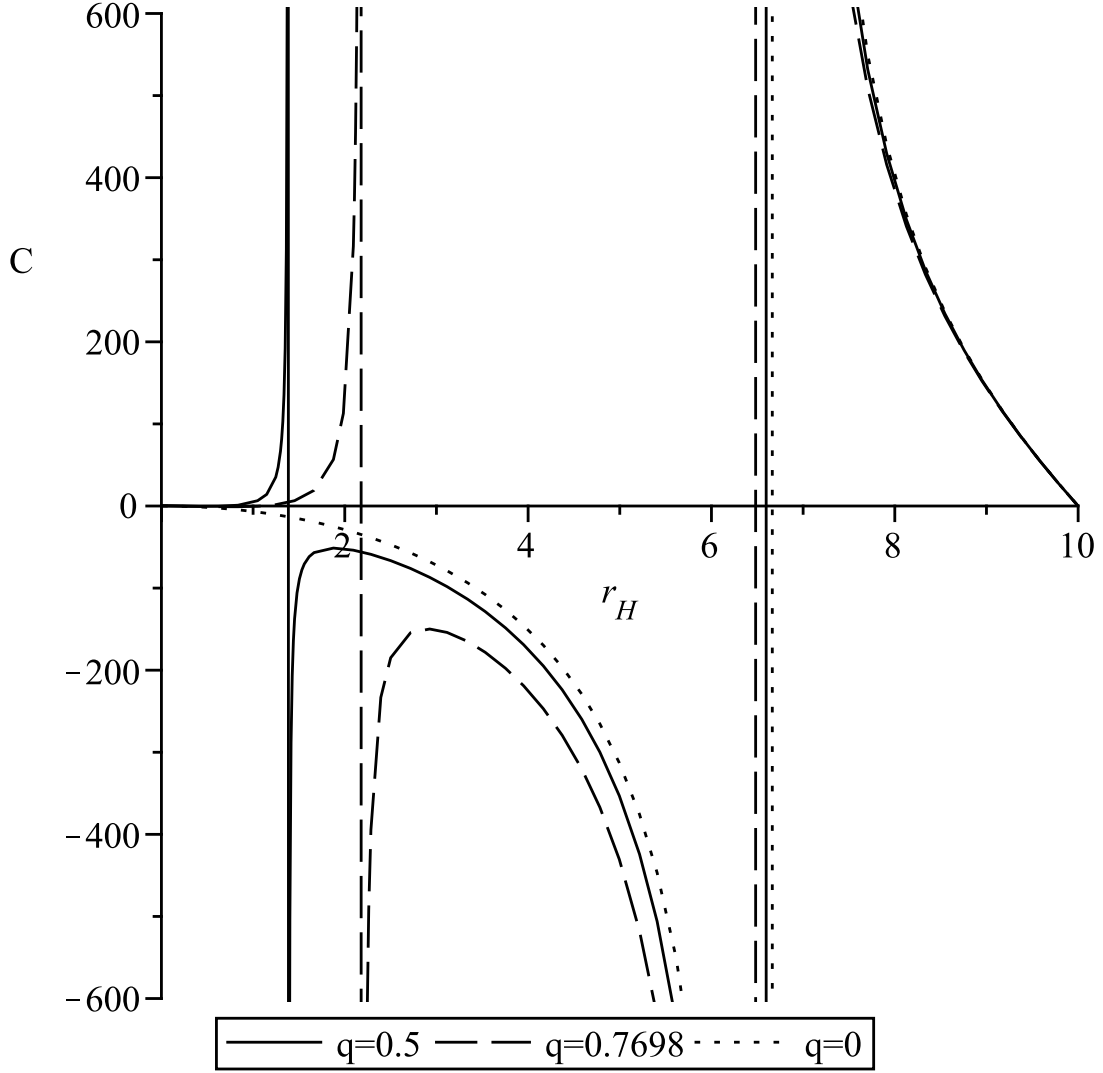}
  \caption{The solid and dotted curves of the dependence of the local temperatures
  on the horizon with $8\pi G\eta^{2}\approx10^{-5}$, $r=10$ and $\psi_{0}=0.02$
  for the Schwarzschild black hole with a global monopole or an $f(R)$ global monopole
respectively.}
\end{figure}

\newpage
\begin{figure}
\setlength{\belowcaptionskip}{10pt} \centering
  \includegraphics[width=15cm]{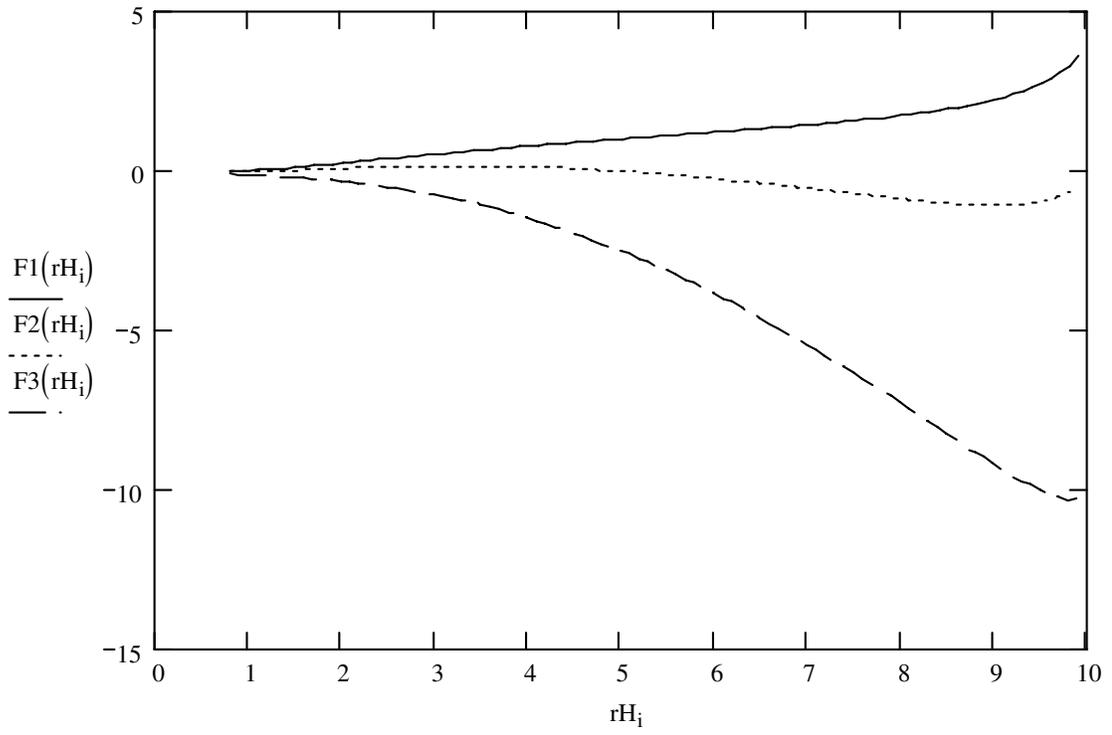}
  \caption{The solid, dotted and dashed curves of the off-shell free energy
  of the Bardeen black holed as a function of the horizon with $q=0.5$
  under the temperature $T=0.015, 0.028, 0.06$ respectively.}
\end{figure}

\newpage
\begin{figure}
\setlength{\belowcaptionskip}{10pt} \centering
  \includegraphics[width=15cm]{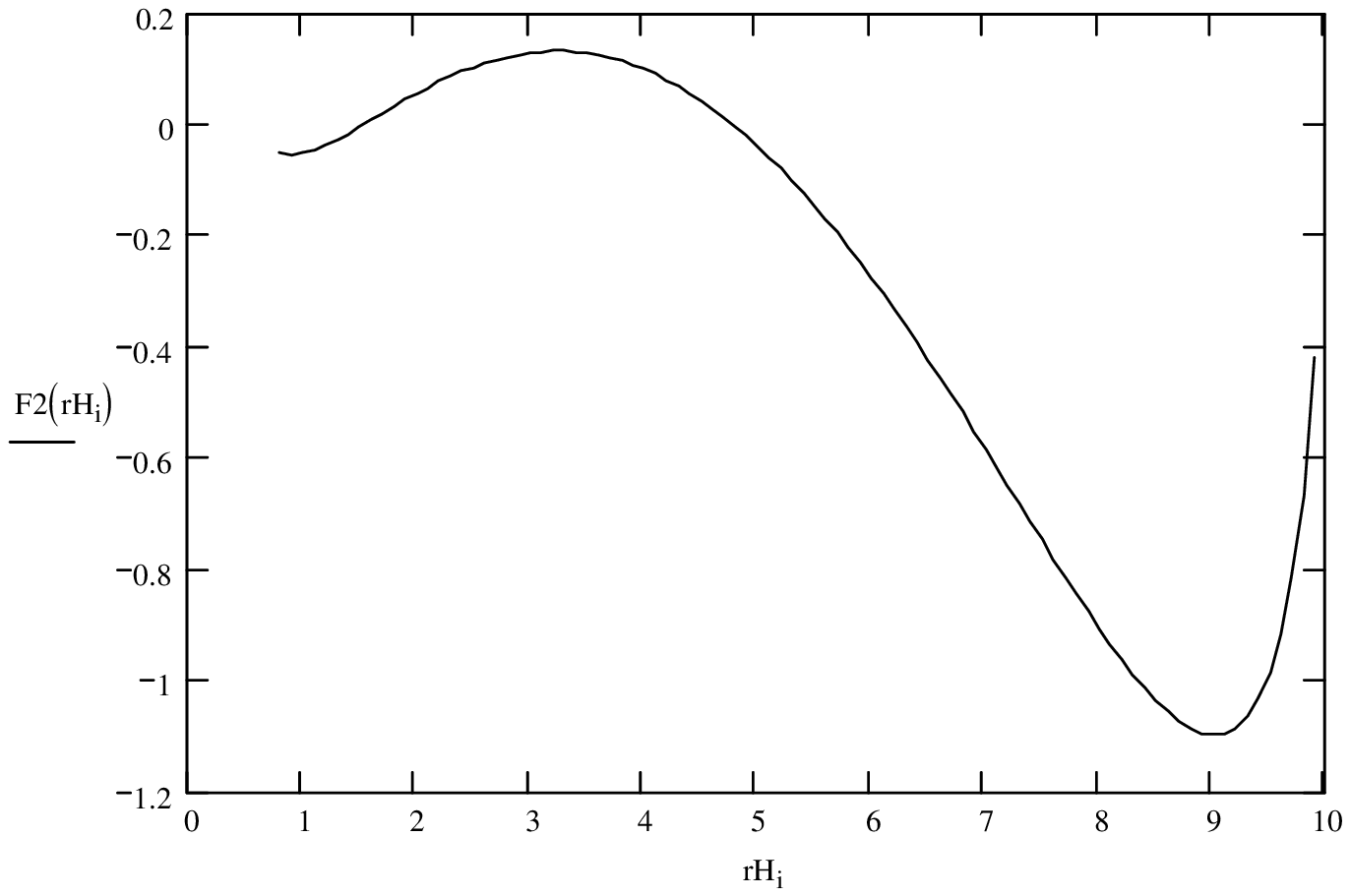}
  \caption{The dependence of the off-shell free energy
  of the Bardeen black holed the horizon with $q=0.5$ and $r=10$
  under the temperature $0.028$.}
\end{figure}


\newpage
\begin{thebibliography}{99}
\bibitem{Bekenstein}J. D. Bekenstein, Lett. Nuovo Cim. 4(1972)737
\bibitem{Bekenstein}J. D. Bekenstein, Phys. Rev. D7(1973)2333
\bibitem{Bekenstein}J. D. Bekenstein, Phys. Rev. D9(1974)3292
\bibitem{Hawking}S. W. Hawking, Commun. Math. Phys. 43(1975)199
\bibitem{Hartle}J. B. Hartle, S. W. Hawking, Phys. Rev.
D13(1976)2188
\bibitem{Gibbons}G. W. Gibbons, S. W. Hawking, Phys. Rev.
D15(1977)2752
\bibitem{Stephens}G. J. Stephens, B. L. Hu, Int. J. Theor. Phys.
40(2001)2183
\bibitem{Kim}W. Kim, E. J. Son, M. Yoon, JHEP 0804(2008)042
\bibitem{Cai}R. G. Cai, L. M. Cao, N. Ohta, JHEP 1004(2010)082
\bibitem{Kim}W. Kim, Y. Kim, Phys. Lett. B718(2012)687
\bibitem{Myung}Y. S. Myung, Y. Kim, Y. Park, Phys. Rev.
D78(2008)084002
\bibitem{Lala}A. Lala, D. Roychowdhury, Phys. Rev. D86(2012)084027
\bibitem{Bardeen}J. Bardeen, Proceedings of GR5, Tiflis, U. S. S.
R
\bibitem{Borde}A. Borde, Phys. Rev. D50(1994)3692
\bibitem{Borde}A. Borde, Phys. Rev. D55(1997)7615
\bibitem{Ayon-Beato}E. Ayon-Beato, A. Garcia, Phys. Lett.
B493(2000)149
\bibitem{Ayon-Beato}E. Ayon-Beato, A. Garcia, Phys. Lett.
B464(1999)25
\bibitem{Ayon-Beato}E. Ayon-Beato, A. Garcia, Gen. Relativ. Grav.
31(1999)629
\bibitem{Ayon-Beato}E. Ayon-Beato, A. Garcia, Gen. Relativ. Grav.
37(1999)635
\bibitem{Eiroa}E. F. Eiroa, C. M. Sendra, Class. Quantum Grav.
28(2011)085008
\bibitem{Zhou}S. Zhou, J. Chen, Y. Wang, arXiv: 1112.5909
\bibitem{Moreno}C. Moreno, O. Sarbach, Phys. Rev. D67(2003)024028
\bibitem{Fernando}S. Fernando, J. Correa, Phys. Rev.
D86(2012)064039
\bibitem{Sharif}M. Sharif, W. Javed, J. Korean Phys. Soc.
57(2010)217
\bibitem{Sharif}M. Sharif, W. Javed, Can. J. Phys. 89(2011)1027
\bibitem{Tolman}R. C. Tolman, Phys. Rev. 35(1930)904

\end{thebibliography}
\end{document}